\documentclass[review]{elsarticle}

\usepackage{graphicx}
\usepackage{epstopdf}
\usepackage{subcaption}
\usepackage{color}
\usepackage{rotating}

\journal{Journal of \LaTeX\ Templates}

\begin{document}

\begin{frontmatter}



\title{Parametric study of the solar wind interaction with the Hermean magnetosphere for a weak interplanetary magnetic field}


\author[label1,label2]{J. Varela}
\ead{deviriam@gmail.com (telf: +33782822476)}

\author[label2]{F. Pantellini}
\author[label2]{M. Moncuquet}

\address[label1]{AIM DSM/IRFU/SAp, CEA Saclay, France}
\address[label2]{LESIA, Observatoire de Paris, CNRS, UPMC, Universite Paris-Diderot, 5
place Jules Janssen, 92195 Meudon, France}

\begin{abstract}

The aim of this study is to simulate the interaction of the solar wind with the Hermean magnetosphere when the interplanetary magnetic field is weak, performing a parametric study for all the range of hydrodynamic values of the solar wind predicted on Mercury for the ENLIL + GONG WSA + Cone SWRC model: density from $12$ to $180$ cm$^{-3}$, velocity from $200$ to $500$ km/s and temperatures from $2 \cdot 10^4$ to $18 \cdot 10^4$ K, and compare the results with a real MESSENGER orbit as reference case. We use the code PLUTO in spherical coordinates and an asymmetric multipolar expansion for the Hermean magnetic field. The study shows for all simulations a stand off distance larger than the Mercury radius and the presence of close magnetic field lines on the day side of the planet, so the dynamic pressure of the solar wind is not high enough to push the magnetopause on the planet surface if the interplanetary magnetic field is weak. The simulations with large dynamic pressure lead to a large compression of the Hermean magnetic field modifying its topology in the inner magnetosphere as well as the plasma flows from the magnetosheath towards the planet surface.

\end{abstract}

\begin{keyword}
 
94.05.-a, 94.30.vf, 96.30.Dz

\end{keyword}

\end{frontmatter}


\section{Introduction}
\label{Introduction}

The analysis of MESSENGER magnetometer data revealed that the Hermean magnetic field can be described as a multipolar expansion \cite{2012JGRE..117.0L12A} with a dipolar moment of $195nT*R^{3}_{M}$ (with $R_{M}$ the planetary radius) as well as the relative small proportion between the strength of the interplanetary magnetic field (IMF) and the Hermean magnetic field $\alpha = B_{sw}//B_{M}$ \cite{2009JGRA..11410101B,2011PandSS...59.2066B}. The range of $\alpha$ values oscillates from 0.3 during a coronal mass ejection ($B_{sw} \approx 65 $ nT) to 0.04 for a period of low magnetic activity of the Sun ($B_{sw} \approx 8 $ nT) \cite{2011Sci...333.1859A,JGRE:JGRE3136}. There is a large variety of magnetosphere configurations due to the interaction between the interplanetary and the Hermean magnetic field and it is particularly relevant the effect of the magnetic reconnection in the bow shock (BS) stand off distance \cite{2007SSRv..132..529F}.

The IMF is not the only free parameter in the solar wind (SW) configuration, there is also a range of possible values for the SW density, velocity and temperature \cite{2009JGRA..11410101B,2011PandSS...59.2066B}. MESSENGER instruments don't measure the hydrodynamic properties of the SW, so the present modeling efforts only includes values obtained by numerical models as the ENLIL + GONG WSA + Cone SWRC \cite{2013JGRA..118...45B}.

The aim of this study is to clarify the effect of the SW hydrodynamic parameters in the Hermean magnetosphere, calculating the location of the stand off distance, the shape of the BS, the regions with strong inflow, open magnetic field lines and mass deposition on the planet surface, the Hermean magnetic field topology and the properties of the plasma stream that links the magnetosheath with the planet surface.

We perform a parametric study in all the range of realistic values predicted by the ENLIL + GONG WSA + Cone SWRC model: density $12 - 180$ cm$^{-3}$, velocity $200 - 500$ km/s and temperature $2 \cdot 10^{4} - 18 \cdot 10^{4}$ K \cite{ODSTRCIL2003497,SWE:SWE449}. To minimize the effect of the IMF in the Hermean magnetosphere we select as reference case a SW configuration with a weak IMF of $7.28$ nT. The SW configuration during coronal mass ejections are excluded from this analysis because these events are characterized by plasma velocity larger than $500$ km/s and a strong IMF mainly oriented in the Southward direction (strong reconnection case). To isolate the effect of a hydrodynamic SW parameters we perform the simulations fixing the other parameters to the same values than the reference case.

We use the MHD version of the single fluid code PLUTO in the ideal and inviscid limit for spherical 3D coordinates \cite{2007ApJS..170..228M}. The Northward displacement of the Hermean magnetic field is represented by a multipolar expansion \cite{2008Sci...321...82A}. The IMF values are obtained from MESSENGER magnetometer data.

This study is called to complement observational studies of the magnetosheath plasma depletion \cite{2013JGRA..118.7181G,2013AGUFMSM24A..03D}, including a comprehensive analysis of the SW hydrodynamic parameters effects, an extension of previous theoretical studies devoted to simulate the global structures of the Hermean magnetosphere using MHD \cite{2008Icar..195....1K} and Hybrid \cite{2007AGUFMSM53C1412T,2010Icar..209...11T} numerical models.

This paper is structured as follows. Section 2, we do a model description including the code features, axisymmetric Hermean magnetic field, boundary and initial. Section 3, we describe the results for the reference case. Section 4, we show the results of the parametric study for the density, velocity and temperature and the effect of the SW hydrodynamic parameters on the plasma flows towards the planet surface. Section 5, conclusion, discussion and context of the study.

\section{Numerical model}
\label{Model}

We use the MHD version of the code PLUTO in the ideal and inviscid limit for a single polytrophic  fluid in 3D spherical coordinates. The code is freely available online \cite{2007ApJS..170..228M}.

The simulation domain is confined within two spherical shells, representing the inner (planet) and outer (solar wind) boundaries of the system. Between the inner shell and the planet surface (at radius unity in the domain) there is a "soft coupling region" where special conditions apply (defined in the next section).The shells are at $0.6 R_{M}$ and $12 R_{M}$ ($R_{M}$ is the Mercury radius).

The conservative form of the equations are integrated using a Harten, Lax, Van Leer approximate Riemann solver (hll) associated with a diffusive limiter (minmod). The divergence of the magnetic field is ensured by a mixed hyperbolic/parabolic divergence cleaning technique (DIV CLEANING) \cite{2002JCoPh.175..645D}.

The grid points are $196$ radial points, $48$ in the polar angle $\theta$ and $96$ in the azimuthal angle $\phi$ (the grid poles correspond to the magnetic poles).

The planetary magnetic field is an axisymmetric model with the magnetic potential $\Psi$ expanded in dipolar, quadrupolar, octupolar and 16-polar terms \cite{2013AGUFMSM24A..03D}:

$$ \Psi (r,\theta) = R_{M}\sum^{4}_{l=1} (\frac{R_{M}}{r})^{l+1} g_{l0} P_{l}(cos\theta) $$  
The current free magnetic field is $B_{M} = -\nabla \Psi $. $r$ is the distance to the planet center and $\theta$ the polar angle. The Legendre polynomials of the magnetic potential are:

$$ P_{1}(x) = x $$
$$ P_{2}(x) = \frac{1}{2} (3x^2 - 1) $$
$$ P_{3}(x) = \frac{1}{2} (5x^3 - 3x) $$
$$ P_{4}(x) = \frac{1}{2} (35x^4 - 30x^2 + 3) $$
the numerical coefficients $g_{l0}$ taken from Anderson et al. 2012 are summarized in the Table 1.

\begin{table}[h]
\centering
\begin{tabular}{c | c c c c}
coeff & $g_{01}$(nT) & $g_{02}/g_{01}$ & $g_{03}/g_{01}$ & $g_{04}/g_{01}$  \\ \hline
 & $-182$ & $0.4096$ & $0.1265$ & $0.0301$ \\
\end{tabular}
\caption{Multipolar coefficients $g_{l0}$ for Mercury's internal field.}
\end{table}

The simulation frame is such that the z-axis is given by the planetary magnetic axis pointing to the magnetic North pole and the Sun is located in the XZ plane with $x_{sun} > 0$. The y-axis completes the right-handed system.

\subsection{Boundary conditions and initial conditions}

The outer boundary is divided in two regions, the upstream part (left in the figure) where the solar wind parameters are fixed and the downstream part (right in the figure) where we consider the null derivative condition $\frac{\partial}{\partial r} = 0$ for all fields. In the inner boundary the value of the intrinsic magnetic field of Mercury and the density are fixed. In the soft coupling region the velocity is smoothly reduced to zero in the inner boundary, the magnetic field and the velocity are parallel, and the profiles of the density is adjusted to keep the Alfven velocity constant $v_{A} = B / \sqrt{\mu_{0}\rho} = 25$ km/s with $\rho = nm_{p}$ the mass density, $n$ the particle number, $m_{p}$ the proton mass and $\mu_{0}$ the vacuum magnetic permeability. In the initial conditions we define a cone in the night side of the planet with zero velocity and low density centered in the planet. The IMF is cut off at $2 R_{M}$.

\section{Reference case}
\label{Reference case}

In this section we perform the analysis of the reference simulation showing the global magnetosphere strucutres and the flows toward the planet surface. The solar wind parameters in the reference simulation are summarized in the Table 2. We assume a fully ionized proton electron plasma, the sound speed is defined as $c_{s} = \sqrt {\gamma*p/\rho} $ (with $p$ the total electron and proton pressure), and the sonic Mach number as $M_{s} = v/c_{s}$ with $v$ the velocity. 

\begin{table}[h]
\centering
\begin{tabular}{c | c c c c c c c c}
Date & B field (nT) & n (cm$^{-3}$) & T (K)  & $\beta$ & $V$ (km/s) & $M_{s}$  \\ \hline
2011/10/10 & $(4,1,6)$ & $60$ & $58000$ & $2.27$ & $250$ & $6.25$ \\
\end{tabular}
\caption{Reference simulation parameters}
\end{table}

In the Figure 1 we compare the MESSENGER data (black line) and the simulation magnetic field (red line) along the satellite trajectory. The location of the main structures of the magnetosphere (bow shock and magnetopause) along the trajectory shows a fair agreement between the MESSENGER data and the simulation. The magnetic field components show the same rotations and profile flattening, pointing out that the model can reproduce the global structures and the Hermean magnetic field topology.

\begin{figure}[h]
\centering
\includegraphics[width=0.5\textwidth]{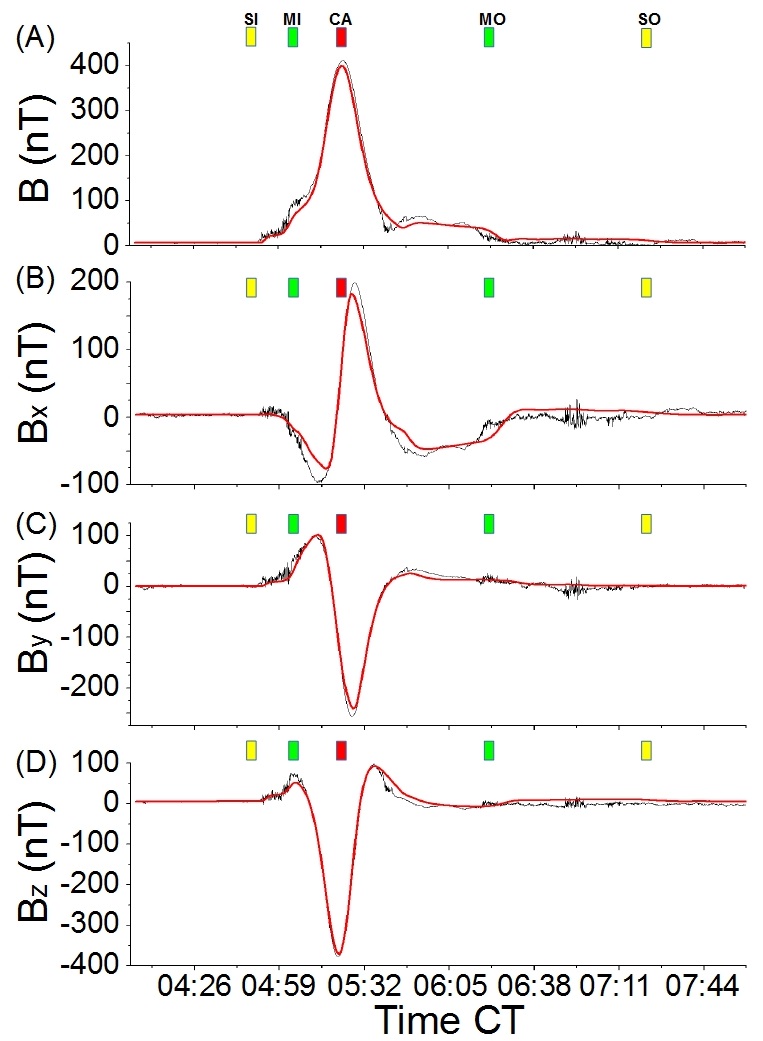}
\caption{Comparison of MESSENGER data (black line) with the simulation magnetic field (red line) along the satellite trajectory. The location of the satellite encounter with the BS (SI and SO), the magnetopause (MI and MO) and the closest approach (CA) are included in the graphs. The data corresponds to the orbit of the day 10 of October of 2011.}
\end{figure}

Figure 2 shows the density distribution in a polar (A) and an equatorial (B) cut. The magnetosheath is identified as the region of high density between the BS and the magnetopause (the magnetopause is defined as a global magnetosphere structure where the planetary magnetic field begins to dominate over the IMF). The inner magnetosphere is the region of low density beyond the magnetopause. There are close magnetic field lines on the day side of the planet, with the BS equatorial stand off distance located $0.87 \cdot R_{M}$ further form the planet surface and the magnetopause $0.6 \cdot R_{M}$. There is a magnetosphere structure (defined in the text as plasma stream) that links the back of the magnetosheath with the planet surface in the cusp region (transition between the open/closed magnetic field lines near the planet poles). The graph C shows the region with inflow/outflow (blue/red) and open magnetic field lines (cyan dots) on the planet surface. The largest inflows are observed at the South Hemisphere dayside (at middle-high latitudes) although there is a local maximum too at the North Hemisphere near the pole. The local maximum of the inflow at both Hemispheres is correlated with the presence of the plasma stream and a region of open magnetic field lines. The mass deposition at the North (D) and South (E) Hemispheres indicates stronger plasma flows at the South Hemisphere. The mass deposition at the North Hemisphere is located mainly near the poles while at the South Hemisphere the deposition region is extended from the poles to middle latitudes.

\begin{figure}[h]
\centering
\includegraphics[width=0.5\textwidth]{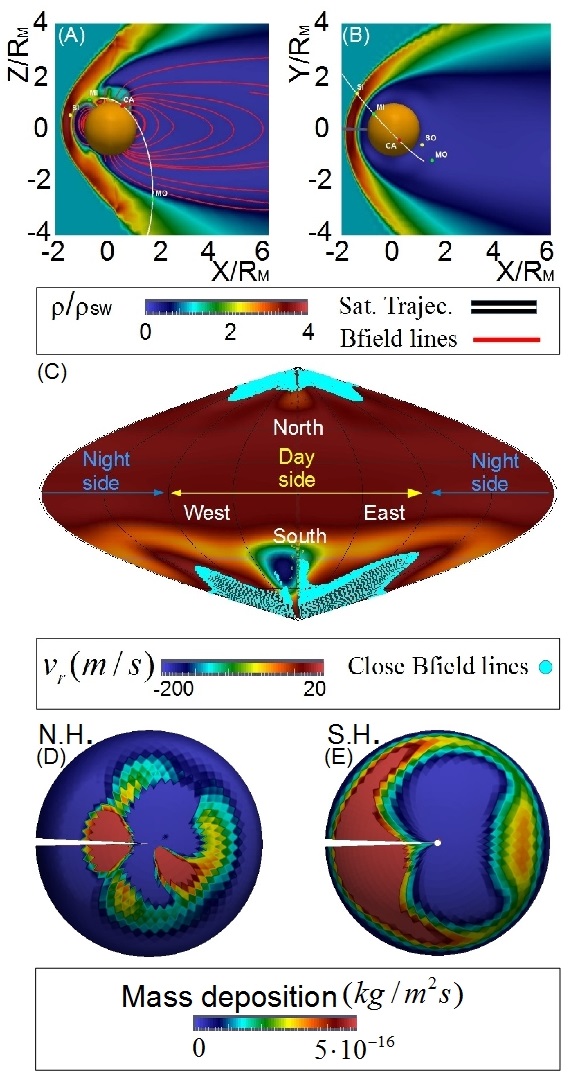}
\caption{Density distribution for a polar (A) and equatorial (B) cuts. We include the points along the trajectory where the satellite reaches the BS (SI and SO), magnetopause (MI and MO) and closest approach (CA). The red lines show the magnetic field lines inside the magnetosphere. (C)  Sinusoidal (Sanson-Flamsteed) projection of the inflow/outflow (blue/red) and open magnetic field lines regions (cyan dots) on the planet surface. Mass deposition at the North (D) and South (E) Hemisphere.}
\end{figure}

\section{Parametric studies}
\label{Parametric studies}

In the following we study the effect of different configuration of the SW hydrodynamic parameters in the magnetospheric global structures and plasma flows towards the planet surface compared with the reference case.

\subsection{Magnetosphere structure}

The figure 3 shows the BS and magnetopause stand off distance at planes rotated $\theta = 0^{o}$, $30^{o}$, $60^{o}$ and $90^{o}$ respect to the equatorial plane at the North Hemisphere day side for the parametric study of the density (A), velocity (B) and temperature (C). All the other simulation parameters are the same than in the reference case (only one parameter is modified at the time). The main magnetospheric structures are located closer to the planet as the density or the velocity increase because the dynamic pressure ($q = \rho v^2/2$) of the SW increases, as can be expected from the balance between the SW dynamic pressure and the magnetic pressure of the Hermean magnetic field:

$$ \frac{R_{MP}}{R_{M}} = \left(\frac{B^{2}}{m_{p}n\mu_{0}v^{2}}\right)^{1/6} $$
where $R_{MP}$ is the location of the magnetopause and $B$ is the dipolar Hermean magnetic field module. There is no dependency of the magnetopause location with the temperature but a hotter SW leads to a reduction of the sonic Mach number due to the increase of the sound speed $c_{s}^2 = \gamma T k_{B} / m_{p} $, with $T$ the temperature and $k_{B}$ the Boltzmann constant. The drop of the Mach number is correlated with an expansion of the BS, located further from the planet as the temperature increases, but no dependency between the magnetopause position and the temperature is observed. The magnetopause is in all simulations is over the Hermean surface so there are closed magnetic field lines on the day side of the planet. For all the simulations the theoretical prediction of the magnetopause location in the equatorial plane (gray dashed line) is a $7.6 \%$ smaller in average (compared with the yellow dashed line), indicating that even for a low value of the IMF module the slightly Northward orientation of the IMF leads to an enhancement of the Hermean magnetic field in the equatorial region, and the magnetopause is located further from the planet.

\begin{figure}[h]
\centering
\includegraphics[width=1.0\textwidth]{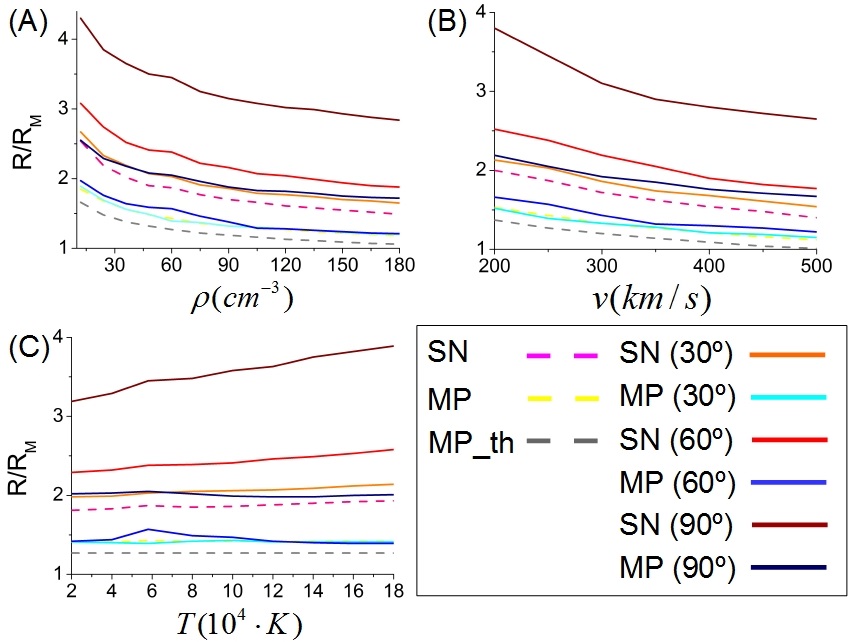}
\caption{BS and magnetopause stand off distance at planes rotated $\theta = 0^{o}$, $30^{o}$, $60^{o}$ and $90^{o}$ respect the equatorial plane at the North Hemisphere day side. The graph (A) shows the results for the parametric study of the density, graph (B) for the velocity and graph (C) for the temperature (all the other simulation parameters are the same than in the reference case, only one parameter is modified at the time). The gray dashed line indicates the theoretical position of the magnetopause in the equatorial plane from the balance between the solar wind dynamic pressure and the magnetic pressure of the Hermean magnetic field.}
\end{figure}

Figure 4 shows the effect of the SW hydrodynamic parameters in the Hermean magnetic field topology, plotting the magnetic field module and components at planes rotated $\theta = 0^{o}$ (A-D), $30^{o}$ (E-H) and $60^{o}$ (I-M) respect to the equatorial plane at the North Hemisphere for the simulations with $\rho = 12$ cm$^{-3}$, $\rho = 180$ cm$^{-3}$, $v = 200$ km/s, $v = 500$ km/s, $T = 2\cdot 10^{4}$ K and $T = 18\cdot 10^{4}$ K (the other values are the same than in the reference case). The module of the magnetic field shows the migration of the BS closer to the planet as the density and the velocity increase, located further if the SW temperature is higher. There are two different regions in the magnetic field module: first the magnetic field module remains constant (along the magnetosheath) and beyond the magnetopause where the magnetic field increases (inner magnetosphere). The inner magnetosphere shows a flattening of the profile between the magnetopause and the closest approach related with the proximity of the reconnection between the Hermean magnetic field and the IMF. If the density and velocity increase or the temperature decreases, the flattening in the inner magnetosphere almost disappears and both regions in the magnetosheath and inner magnetosphere are slender. The magnetic field components show stronger rotations in SW configuration with large dynamic pressure, driven by the compression of the magnetosphere on the day side and the displacement of the cusp to lower latitudes, modifying the Hermean magnetic field topology in the inner magnetosphere. 

\begin{figure}[h]
\centering
\includegraphics[width=1.0\textwidth]{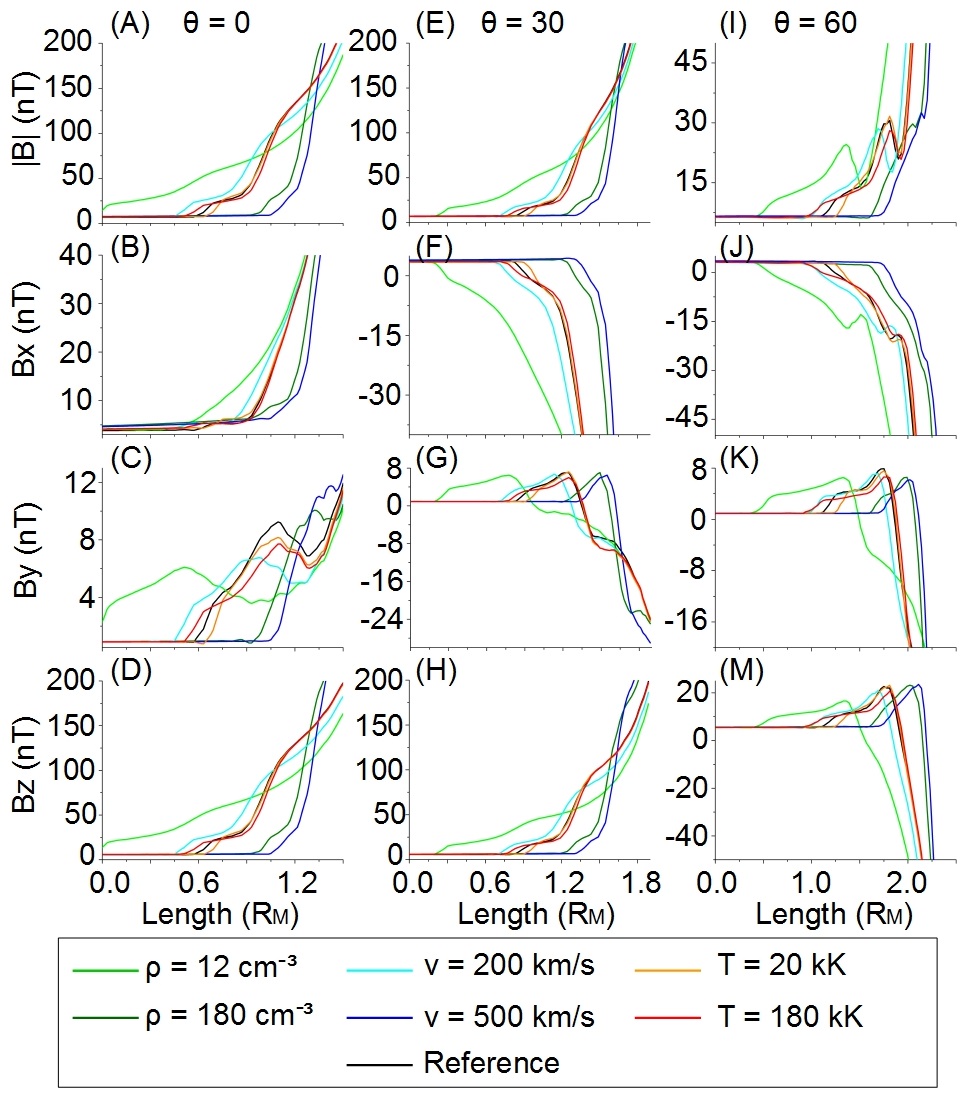}
\caption{Magnetic field module and components at planes with angles of $0^{o}$ (A-D), $30^{o}$ (E-H) and $60^{o}$ (I-M) respect to the equatorial plane at the North Hemisphere for the simulations with $\rho = 12$ cm$^{-3}$, $\rho = 180$ cm$^{-3}$, $v = 200$ km/s, $v = 500$ km/s, $T = 2\cdot 10^{4}$ K and $T = 18\cdot 10^{4}$ K (the other SW parameters are the same than in the reference case).}
\end{figure}

Different SW hydrodynamic parameters induce a dissimilar configuration of the magnetosheath, BS shape and Hermean magnetic field topology, pointing out that the plasma flows towards the planet surface change too. In the next section we analyse the properties of the plasma flows for the different configuration.

\subsection{Plasma precipitation on the planet surface}

The figure 5 shows the density distribution in a polar cut for the simulations with $\rho = 12$ cm$^{-3}$ (A), $\rho = 180$ cm$^{-3}$ (B), $v = 200$ km/s (C), $v = 500$ km/s (D), $T = 2\cdot 10^{4}$ K (E) and $T = 18\cdot 10^{4}$ K (F). The BS is more compressed and the magnetosheath is thinner in the simulations with large dynamic pressure and low temperature. The magnetopause is closer to the planet and the plasma stream structures are shorter, particularly at the South Hemmisphere where the back of the magnetosheath reaches the planet surface for the $\rho = 180$ cm$^{-3}$ and $v = 500$ km/s simulations. The region with closed magnetic field lines on the planet day side decreases as the dynamic pressure increases, mainly located near the equator and at low latitudes of the North Hemisphere.

\begin{figure}[h]
\centering
\includegraphics[width=0.8\textwidth]{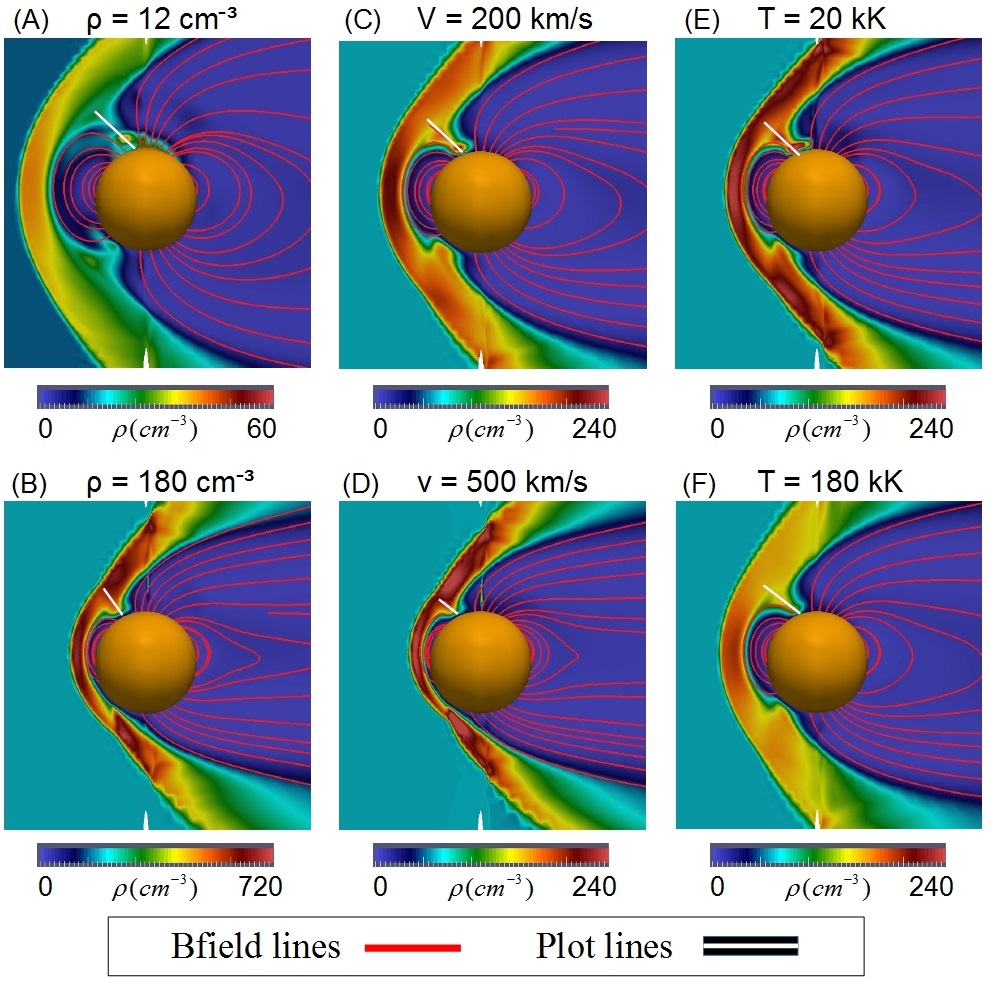}
\caption{Polar cut of the density distributions in the simulations with $\rho = 12$ cm$^{-3}$ (A), $\rho = 180$ cm$^{-3}$ (B), $v = 200$ km/s (C), $v = 500$ km/s (D), $T = 2\cdot 10^{4}$ K (E) and $T = 18\cdot 10^{4}$ K (F). The red lines show the magnetic field lines inside the magnetosphere. The white lines indicate the region plotted in the Figure 6.}
\end{figure}

The figure 6 shows the evolution of the density, temperature, magnetic and velocity field modules as well as the component of the velocity along the plasma stream, from the magnetosheath (left on the graphs) to the planet surface (right on the graphs), for the simulations with $\rho = 12$ cm$^{-3}$ (A), $\rho = 180$ cm$^{-3}$ (D), $v = 200$ km/s (B), $v = 500$ km/s (E), $T = 2\cdot 10^{4}$ K (C) and $T = 18\cdot 10^{4}$ K (F). The plasma stream is originated at the magnetosheath, in the reconnection region between the IMF and the Hermean magnetic field, observed as a local drop of the magnetic field module in the graphs. This region is correlated with a local maximum of the plasma temperature and density as well as a local minimum of the velocity, indicating that the plasma is decelerated, heated and accumulated before precipitate along the open magnetic field line towards the planet surface. During the precipitation the plasma is accelerated, rarefied and cooled. In the simulation with low dynamic pressure (plots A and B) and temperature (plot C), there is a second local maximum of the density nearby the planet surface, correlated with a local minimum of the temperature and a deceleration of the plasma, showing a region of dense and cold plasma accumulated over the North pole before its precipitation on the planet surface. This structure is not observed in the simulations with large dynamic pressure (plots D and E) and temperature (plot F) because the plasma precipitates directly on the planet surface from the magnetosheath. The configuration with the most dense plasma stream is the $\rho = 180$ cm$^{-3}$ simulation while the configuration with the fastest flows along the plasma stream (module and velocity components, in particular the $V_{Z}$) is the $v = 500$ km/s simulation. These SW configurations are the main candidates to drive the largest mass deposition on the planet surface.

\begin{figure}[h]
\centering
\includegraphics[width=0.8\textwidth]{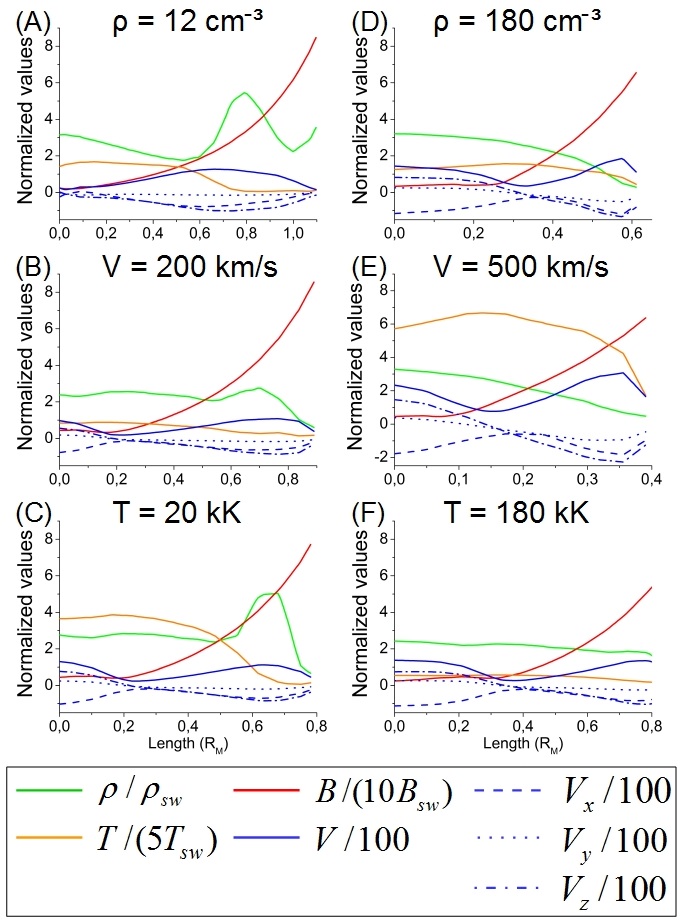}
\caption{Density, temperature, magnetic and velocity field module, and velocity field components along the plasma stream structure (see figure 5) for the simulations with $\rho = 12$ cm$^{-3}$ (A), $\rho = 180$ cm$^{-3}$ (D), $v = 200$ km/s (B), $v = 500$ km/s (E), $T = 2\cdot 10^{4}$ K (C) and $T = 18\cdot 10^{4}$ K (F).}
\end{figure}

Figure 7 shows the regions on the planet surface with inflow/outflow (red/blue) and open magnetic field lines (cyan dots) for the simulations  with $\rho = 12$ cm$^{-3}$ (A), $\rho = 180$ cm$^{-3}$ (D), $v = 200$ km/s (B), $v = 500$ km/s (E), $T = 2\cdot 10^{4}$ K (C) and $T = 18\cdot 10^{4}$ K (F). In the simulations with large dynamic pressure( plots D and E), the flows are enhanced in both Hemispheres due to the magnetosheath compression. The local maximum of the inflow is observed at lower latitudes compared with the reference case because the planet cusp is displaced towards the equator. If we compare the configuration with cold (plot C) and hot (plot F) SW, the inflow regions are similar, slightly enhanced in the case with large temperature. The simulations with low dynamic pressure (plots A and B), show small inflow regions located at high latitudes at both Hemispheres. The regions with open magnetic field lines are wilder in the simulations with larger dynamic pressure, correlated with the local maximum of the inflow, particularly in the South Hemisphere where the BS reaches the planet surface. In the simulations with low dynamic pressure the magnetosphere is more sensitive to the IMF orientation leading to a larger East-West asymmetry of the inflow and open magnetic field lines regions.

\begin{figure}[h]
\centering
\includegraphics[width=1.0\textwidth]{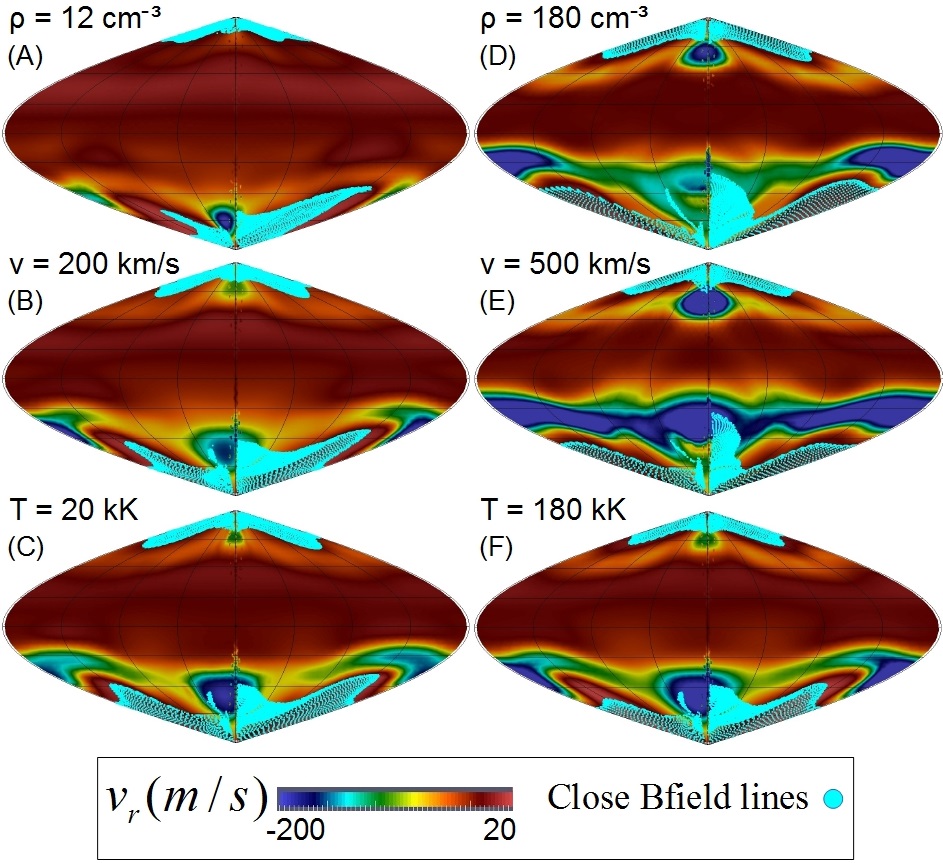}
\caption{Sinusoidal (Sanson-Flamsteed) projection of the inflow/outflow (blue/red) and open magnetic field lines regions (cyan dots) on the planet surface for the simulations with $\rho = 12$ cm$^{-3}$ (A), $\rho = 180$ cm$^{-3}$ (D), $v = 200$ km/s (B), $v = 500$ km/s (E), $T = 2\cdot 10^{4}$ K (C) and $T = 18\cdot 10^{4}$ K (F).}
\end{figure}

Figure 8 shows the mass deposition distribution on the planet surface and the table 3 the integrated value at each Hemisphere for the simulations  with $\rho = 12$ cm$^{-3}$ (A-N, A-S), $\rho = 180$ cm$^{-3}$ (D-N, D-S), $v = 200$ km/s (B-N, B-S), $v = 500$ km/s (E-N, E-S), $T = 2\cdot 10^{4}$ K (C-N, C-S) and $T = 18\cdot 10^{4}$ K (F-N, F-S). The regions of mass deposition are wider for simulations with large dynamic pressure (plots D and E), particularly at the South Hemisphere. The mass deposition region is localized close to the poles on the planet day side in the simulation with low SW density (pplots A-N and A-S) while it is distributed on the night and day side for the slow SW configuration (plots B-N and B-S). The mass deposition for cold and hot SW configurations (plots C and F) is similar, slightly smaller on the night side for a hot plasma. The integrated mass deposition indicates that the main part of the mass deposition takes place at the South Hemisphere, a 50 $\%$ more compared with the North Hemisphere, except in the simulation with slow SW where the mass deposition is similar at both Hemispheres. The mass deposition is more than 20 times larger comparing the configuration with high and low density, 2 times larger comparing the fast and slow SW simulation and almost the same for the hot and cold plasmas configurations. The mass deposition is enhanced at the North Hemisphere for the cold plasma configuration because the magnetosheath is more compressed reinforcing the plasma stream at the North Hemisphere, the opposite scenario than the hot plasma configuration where the mass deposition is enhanced at the South Hemisphere due to the decompression of thmagnetosheath. The simulation with high SW density shows a mass deposition a $15 \%$ larger than the simulation with high velocity.

\begin{figure}[h]
\centering
\includegraphics[width=1.0\textwidth]{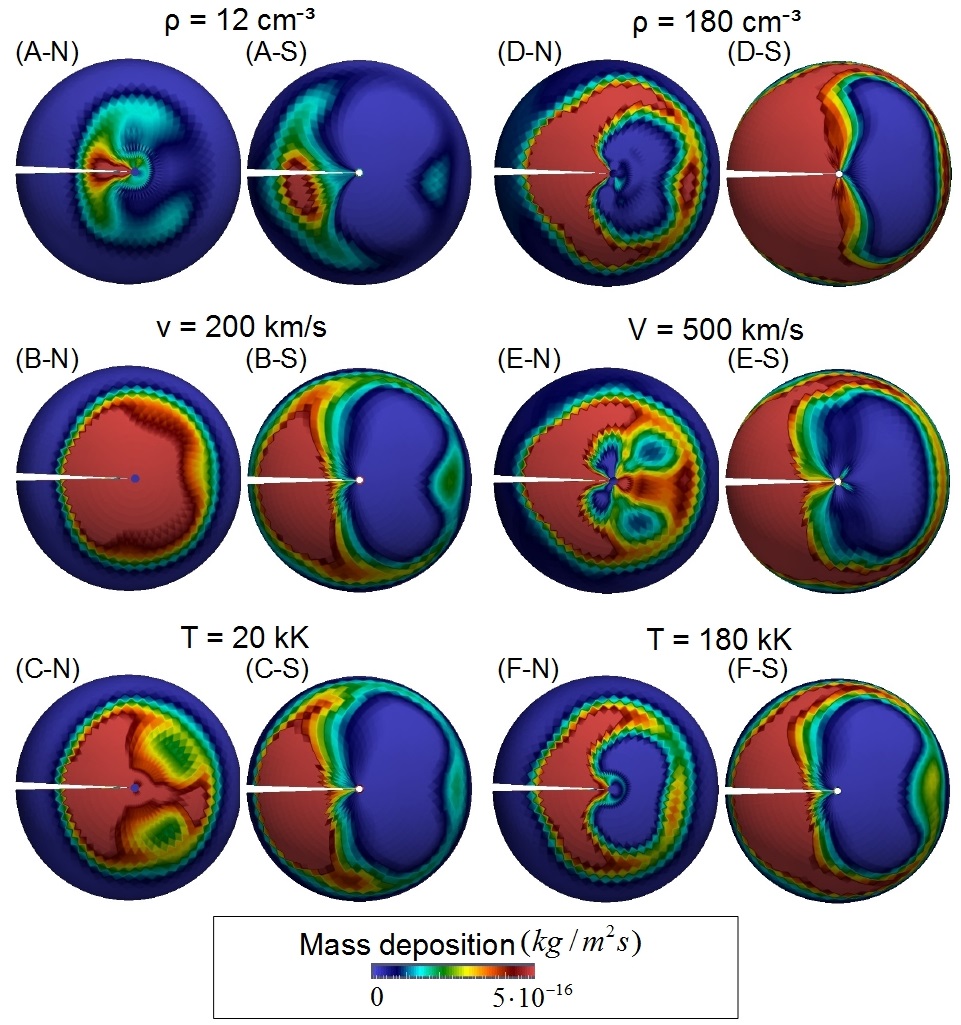}
\caption{Mass deposition at the North and South Hemispheres for the simulations  with $\rho = 12$ cm$^{-3}$ (A-N, A-S), $\rho = 180$ cm$^{-3}$ (D-N, D-S), $v = 200$ km/s (B-N, B-S), $v = 500$ km/s (E-N, E-S), $T = 2\cdot 10^{4}$ K (C-N, C-S) and $T = 18\cdot 10^{4}$ K (F-N, F-S).}
\end{figure}

\begin{table}[h]
\centering
\begin{tabular}{c | c c}
Model & North Hemisphere & South Hemisphere  \\ \hline
Reference & $0.034$ & $0.101$ \\
$\rho = 12$ cm$^{-3}$ & $0.011$ & $0.022$ \\
$\rho = 180$ cm$^{-3}$ & $0.146$ & $0.297$ \\
$v = 200$ km/s & $0.086$ & $0.094$ \\
$v = 500$ km/s & $0.130$ & $0.246$ \\
$T = 2\cdot 10^{4}$ K & $0.077$ & $0.105$ \\
$T = 18\cdot 10^{4}$ K & $0.056$ & $0.124$ \\
\end{tabular}
\caption{Integrated mass deposition at the North and South Hemispheres (kg/s) for the simulations  $\rho = 12$ cm$^{-3}$, $\rho = 180$ cm$^{-3}$, $v = 200$ km/s, $v = 500$ km/s, $T = 2\cdot 10^{4}$ K and $T = 18\cdot 10^{4}$ K.}
\end{table}

\section{Conclusions}
\label{Conclusions}

To perform a parametric study with a weak interplanetary magnetic field illustrates the effect of the solar wind hydrodynamic variables on the Hermean magnetosphere structure, minimizing the distortion driven by the reconnection between the interplanetary and the Hermean magnetic field. The results indicate that the BS never reaches the planet surface in the equator if the SW dynamic pressure is smaller than $6.27 \cdot 10^{-9}$ Pa, the largest dynamic pressure for all the simulations where the SW density is $\rho = 60 $ cm$^{-3}$ and the velocity is $v = 500$ km/s. The forecast of the SW dynamic pressure in Mercury by the ENLIL + GONG WSA + Cone SWRC model usually expects values below $6.27 \cdot 10^{-9}$ Pa, pointing out that the erosion of the Hermean magnetic field by a Southward oriented inteplanetary magnetic field is the main driver of the magnetopause precipitation on the Hermean surface.

Another conclusion of the study is the evolution of the magnetopause and BS stand off distance with the hydrodynamic values, showing that an enhancement of the SW dynamic pressure leads to a more compressed magnetosheath and a more closed magnetosphere (triangular shape of the BS). Hot SW configurations show a decompression of the magnetosheath due to the increase of the SW sound velocity and the drop of the sonic Mach number. The BS front is displaced $0.11 \cdot R_{M}$ comparing the coldest to the hottest solar wind configuration. The theoretical calculation considering only the dynamic pressure of the SW is slightly different than the values obtained in the simulations due to the small but observable effect of the Northward IMF orientation, enough to enhance the Hermean magnetic field in the nose of the bow shock and slightly displace further the magnetopause. 

The simulations with large SW dynamic pressure drives a strong compression of the magnetic field lines on the day side and the magnetotail stretching on the night side, changing the Hermean magnetic field topology of the inner magnetosphere. In consequence, the plasma flows and mass deposition on the planet surface are altered by the different SW configurations.

The inflow and open magnetic field lines regions on the planet surface are wider for configurations with large dynamic pressure. The magnetosheath depletion is more efficient in dense and fast SW configurations. The integrated mass deposition is a $15 \% $ larger in the $\rho = 180$ cm$^{-3}$ simulation than in the $v = 500$ km/s case, even if the dynamic pressure for the high density simulation is a $75 \%$ of the dynamic pressure of the high velocity case. On the other side, the configurations with low SW density leads to integrated mass depositions much lower than the simulation with slow SW, almost 5 times smaller if we compare the simulation with $\rho = 12$ cm$^{-3}$ and $v = 200$ km/s, where the dynamic pressure of the low density simulation is only a $33 \%$ of the low velocity case. The integrated mass deposition is almost the same for hot and cold SW configuration, but the ratio between Hemispheres changes; a $42 \%$ of the total deposition takes place at the North Hemisphere for the $T = 2\cdot 10^{4}$ K simulation versus a $31 \%$ for the $T = 18\cdot 10^{4}$ K case, due to the larger magnetosheath compression in the cold plasma configuration. In summary, there is not a direct correlation between the mass deposition and the dynamic pressure of the SW, it is required a further analysis of the magnetosheath region where the plasma stream is originated to understand the effect of the SW hydrodynamic parameter in the flows towards the Hermean surface.

The plasma stream is originated closer to the planet surface in the simulations with large dynamic pressure and the magnetosheath is slender compared with the low dynamic pressure cases. The influence of the reconnection region covers all the magnetosheath in the large dynamic pressure simulations leading to an enhancement of the plasma precipitation.

The plasma stream is collimated by the Hermean magnetic field in the simulations with low dynamic pressure, leading to small deposition region near the poles. For the high dynamic pressure cases the flows are strong enough to overcome the collimation leading to a spread plasma stream that convers the day and night side of the planet surface.

In the configurations with low dynamic pressure and temperature there is a region near the planet North pole of cold and dense plasma. This structure is not observed in configurations with large SW dynamic pressure because the magnetopause is located too close to the planet surface. In the case of hot SW configurations, the magnetosheath decompression leads to a drop of the plasma precipitation on the North Hemisphere, avoiding a large accumulation of plasma near the pole.

The resolution of the model is not large enough to resolve the plasma depletion layer as a different structure than the magnetosheath, although the simulation conclusion are similar to observational studies showing compatible features for the particles fluxes and magnetosheath depletion \cite{2013AGUFMSM24A..03D,2007SSRv..132..433K}. The numerical resistivity of the code is several orders larger than the real plasma so no magnetic field pile-up is observed on the day side  \cite{2013JGRA..118.7181G}, the reconnection is almost instantaneous, but the simulation can reproduce the important effect of the reconnection in the origin of the plasma stream. The simulation conclusions agrees with the last observations of proton precipitations from the magnetosheath towards the planet surface along the cusp, pointing out that there is not a direct precipitation of the SW at the North pole \cite{2014JGRA..119.6587R}. The present simulations share similar magnetosphere global structures than other simulations performed with different numerical schemes \cite{2007AGUFMSM53C1412T,2010Icar..209...11T}. Present research complements a recent communication of the authors devoted to study the effect of the interplanetary magnetic field orientation on the fluxes toward the Hermean surface \cite{2015PandSS..119..264V}.

\section{Aknowledgments}
The research leading to these results has received funding from the European Commission's Seventh Framework Programme (FP7/2007-2013) under the grant agreement SHOCK (project number 284515). The MESSENGER magnetometer data set was obtained from the NASA Planetary Data System (PDS) and the values of the solar wind hydrodynamic parameters from the NASA Integrated Space Weather Analysis System.





\section*{References}

\bibliography{mybibfile}

\end{document}